\documentclass[12pt]{iopart}
\usepackage{graphicx,harvard,amssymb}

\newcommand{\bra}[1]{\mathop{\left\langle #1 \right|}\nolimits}
\newcommand{\ket}[1]{\mathop{\left| #1 \right\rangle}\nolimits}

\newcommand{\braket}[2]{\mathop{\left\langle #1 \right|\left. #2\right\rangle}\nolimits}

\begin{document}

\title{Coherent State Quantum Key Distribution with Multi Letter Phase-Shift Keying}
\author{Denis Sych and Gerd Leuchs}
\address{Max Planck Institut f\"ur die Physik des Lichts, G\"unther--Scharowsky--Strasse 1 / Bau24,
D-91058 Erlangen, Germany}
\address{Institut f\"ur Optik, Information und Photonik, Universit\"at Erlangen--N\"urnberg, Staudtstrasse 7 / B2, 91058 Erlangen, Germany}
\date{\today}
\begin{abstract}
We present a protocol for quantum key distribution using discrete modulation of coherent states of light. Information is encoded in the variable phase of coherent states which can be chosen from a regular discrete set ranging from binary to continuous modulation, similar to phase--shift--keying in classical communication. Information is decoded by simultaneous homodyne measurement of both quadratures and requires no active choice of basis. The protocol utilizes either direct or reverse reconciliation, both with and without postselection. We analyze the security of the protocol and show how to enhance it by the optimal choice of all variable parameters of the quantum signal.
\end{abstract}
\maketitle

\section{Introduction}

Quantum key distribution (QKD) is a procedure of information exchange between two parties, the sender Alice and the receiver Bob, which allows to distribute absolutely secure data between them \cite{Gisin:02RMP,Dusek:06,Scarani:09}. The distinctive part of QKD with respect to classical communication is the use of a quantum information channel, where the signal is protected from unauthorized duplication \cite{Wooters:82,Dieks:82,Bennett:84,Ekert:91}. Mathematically, the information signal can be described by discrete variables (DV) or by continuous variables (CV) \cite{Cerf:07book}, although physically the signal can be of various kinds: single photons \cite{Bennett:84}, weak coherent pulses \cite{Bennett:92}, squeezed states \cite{Ralph:99,Hillery:00} and other systems where a signal possesses essentially quantum properties. 

A universal characteristic that can be compared for different QKD protocols (apart from their experimental realizations) is the secret key generation rate (shortly, {\em key rate})~--- i.e. the average amount of secure information per elementary transmission (e.g. per light pulse). In order to transmit higher total amount of secure data one can increase the pulse repetition rate or increase the key rate. The first way is limited mainly by experimental techniques, while the second one is defined by mathematical properties of the given QKD protocol.

In the case of DV QKD, it has been shown that extensions of the standard four--letter BB84 protocol \cite{Bennett:84} to higher number of letters in the alphabet can improve performance in terms of higher critical error rate or longer communication lines \cite{Brus:98,Bech:99,Sych:04,Sych:05}.

In the case of CV QKD, the first protocols were based on Gaussian alphabets consisting of either coherent or squeezed states. The common weak point of these protocols is high sensitivity to losses in the quantum communication channel, which initially was believed to lead to the so--called ``3 dB loss limit'': the key rate is equal to zero when channel losses are higher than $50\%$. After the invention of postselection \cite{Silb:02} and reverse reconciliation \cite{Gross:03}, this limit was overcome, and the key rate was substantially improved. Another interesting option for improving the key rate is to consider discrete alphabets \cite{Namiki:03,Heid:06} instead of Gaussian ones. Protocols with Gaussian alphabets have an advantage of simpler security analysis, whilst the protocols based on discrete modulation are easier to realize in practice.

In this work we address the question how one can further increase the key rate of CV QKD with discrete modulation by varying the geometry of the quantum alphabet. Having in mind the idea of improving the properties of DV QKD by use of more symmetric alphabets with higher number of letters \cite{Sych:05}, we present a new CV QKD protocol which generalizes previous ones with discrete modulation \cite{Namiki:03,Lorenz:04,Heid:06}. Namely, the protocol employs an alphabet with $N$ coherent states $\ket{\alpha_k}=\ket{a e^{i\frac{2\pi}{N}k}}$ which have relative phases $\frac{2\pi}{N}k$ and a fixed amplitude $a$. In classical communication, this type of encoding is known as phase--shift--keying (PSK). We perform a security analysis of the proposed protocol for lossy but noiseless quantum channels, providing full optimization of all parameters of the protocol, and show how the number of letters affects the key rate.

\section{Description of the protocol}

An elementary information transmission in the multi letter PSK protocol is as follows:
\begin{itemize}
\item  The sender (Alice) chooses a random equiprobable number $k=1\ldots N$ and sends the respective coherent state $\displaystyle\ket{\alpha_k}=\ket{a e^{i\frac{2\pi}{N}k}}$.
\item The receiver (Bob) measures the state by splitting the signal at a 50/50 beam splitter and measuring two conjugate quadratures $\hat x$ and $\hat p$ at the output ports by homodyning each signal~--- such a scheme is called heterodyne measurement \cite{Lorenz:04,Weed:04}, where the two conjugate measurements can be separated in time or space. The results of the measurements are $\beta_x$ and $\beta_p$, which we write as a pure coherent state $\ket{\beta}=\ket{\beta_x+i\beta_p}$.
\item Bob assigns a classical number $l$ to the measured state $\ket{\beta}$ by finding a state $\ket{\alpha_l}$ which is the closest alphabet's state to the state $\ket{\beta}$: $|\braket{\alpha_l}{\beta}|^2=\max\limits_{n}|\braket{\alpha_n}{\beta}|^2$. Generally speaking, this classical value $l$ decoded by Bob can be different from the initial value $k$ sent by Alice because of intrinsic quantum uncertainty even in the absence of eavesdropping or any channel noise.
\end{itemize}
The elementary information transmission from Alice to Bob can be schematically shown as a ``$k\rightarrow l$'' channel:
\begin{equation}\label{ABflow}
k\stackrel{encoding}{\longrightarrow}\ket{\alpha_k}\stackrel{measurement}{\longrightarrow}\ket{\beta}\stackrel{decoding}{\longrightarrow} l.
\end{equation}

As long as the considered quantum alphabet has a regular $2\pi/N$ phase--shift symmetry, and all states have an equal probability to be sent, all channel inputs and outputs (\ref{ABflow}) are equiprobable. We note, that there is no active choice of measurement basis in our protocol, therefore there is no basis reconciliation needed. All transmissions contribute to the total secure key, and nothing is discarded unlike in previous protocols with discrete modulation and homodyne detection \cite{Namiki:06}.

After the measurement, Bob can (but not necessarily has to) use the postselection idea \cite{Silb:02}, when he decides whether to keep the transmission depending on the value $\beta$. The elementary transmission (\ref{ABflow}), possibly followed by postselection, is repeated until Alice and Bob collect enough data to perform classical error correction and privacy amplification procedures \cite{Gisin:02RMP}. The resulting data is a secret key.

The amplitude $a$ and the number of letters $N$ can be flexibly adjusted for a given information channel. The exact optimization for a given channel transmittance will be discussed later. These parameters are supposed to be publicly known, particularly by the potential eavesdropper. We note that the number of states can be arbitrary large. In the limit of infinity $N\rightarrow\infty$ we have a continuous phase modulation. Thus our protocol can be viewed as a smooth transition between discrete \cite{Namiki:03,Lorenz:04,Heid:06} and continuous \cite{Silb:02,Gross:02,Weed:04} modulation of CV.

\section{Security analysis}

We investigate the security of our protocol assuming there is no excess noise in the quantum channel. In this case the best possible attack is the beam splitting attack \cite{Heid:06}. We allow Eve to have unlimited access to all the losses, as she could replace the real lossy information channel with an ideal lossless one.

The beam splitting transformation is $\ket{\alpha_k}_A\rightarrow\ket{\beta_k}_B\otimes\ket{\epsilon_k}_E$, where Alice's initial state $\ket{\alpha_k}$ is split to Bob's state $\ket{\beta_k}$ and Eve's state $\ket{\epsilon_k}$:
\begin{equation}\label{bestates}
	\ket{\beta_k}=\ket{\sqrt\eta\alpha_k},\quad\ket{\epsilon_k}=\ket{\sqrt{1-\eta}\alpha_k},
\end{equation}
and $\eta$ is the channel transmittance. 

In this beam splitting scenario, Eve does not introduce any excess noise on Bob's side, whereas in any other better eavesdropping strategy Eve necessarily does. For example, if Eve would make an intercept--resend attack, then she adds at least one unit of shot noise. Afterwards, she can attenuate the signal, and the excess noise will be proportionally reduced. As a remark on the side, we see in this way, that the maximum tolerable excess noise cannot exceed that of the intercept--resend strategy, i.e. cannot be higher then channel transmittance $\eta$. 

In real communication lines, such as optical fibres or free space, excess noise (typically, about $1\%$ of shot noise \cite{Lorenz:04,Lorenz:06,Elser:08}) is introduced mainly by imperfections of the experimental setup, and there is almost no measurable excess noise due to the channel itself. If the absence of channel excess noise is experimentally verified, then the eavesdropping strategy based on beam splitting is the best possible attack, at least for the values of the channel transmittance $\eta\gg0.01$.

\subsection{Information between Alice and Bob}

The amount of classical mutual Shannon information $I_{AB}$ transmitted from Alice to Bob via the channel (\ref{ABflow}) is equal to the difference of {\em a priori} (before measurement) and {\em a posteriori} (after  measurement) entropies \cite{Shan:48}. Before any measurement, all channel outcomes are equiprobable for Bob, so his {\em a priori} entropy $H_{Bob}^{prior}$ is the unconditional ``pure guess'' entropy equal to $\log_2 N$ bit per transmission. 

The conditional probability density to measure the state $\ket{\beta}$ when a state $\ket{\alpha_k}$ has been sent is $p(\beta|k)\sim |\braket{\beta_k}{\beta}|^2\sim e^{-|\beta-\beta_k|^2}.$ The total unconditional probability density to measure a state $\ket{\beta}$ is $p(\beta)=\frac{1}{N}\sum\limits_{k=1}^{N}p(\beta|k)$.
Its normalization $\int p(\beta)d\beta=1$  also yields the normalization of  $p(\beta|k)=\frac{1}{\pi}e^{-|\beta-\beta_k|^2}.$

After the measurement of $\ket{\beta}$, the probability $p_l(\beta)$ that the state $\ket{\alpha_l}$ was initially sent is
\begin{equation}\label{pl}
 p_l(\beta)=\frac{p(\beta|l)}{N p(\beta)}=\frac{1}{\pi N p(\beta)}e^{-|\beta-\beta_l|^2}.
\end{equation}
As we discussed above, the measured state $\ket{\beta}$ is decoded by Bob to a classical value $l$ such that the state $\ket{\alpha_l}$ is the closest alphabet's state to the measured state $\ket{\beta}$. Corresponding regions in the phase space are shown by different shades of grey in Fig.~\ref{5lett}. In the case when $l=k$ the value (\ref{pl}) is the probability of decoding the correct result, otherwise (\ref{pl}) is the error probability of decoding a wrong result $l\neq k$.

Bob's {\em a posteriori} entropy $H_{Bob}^{post}$ is the Shannon entropy of the total probability distribution $P(\beta)=\{p_1(\beta),p_2(\beta),\ldots,p_N(\beta)\}$ conditioned on the measured state $\ket{\beta}$:
\begin{equation}\label{ShanEnt}
 H_{Bob}^{post}[P(\beta)]=-\sum\limits_{k=1}^{N}p_k(\beta)\log_2 p_k(\beta).
\end{equation}

Finally, the amount of classical information transmitted from Alice to Bob via the channel (\ref{ABflow}):
\begin{equation}\label{Iab}
	I_{AB}=\int p(\beta)I_{AB}(\beta)d\beta,
\end{equation}
where $I_{AB}(\beta)=\log_2 N+\sum\limits_{k=1}^{N}p_k(\beta)\log_2 p_k(\beta)$.

\subsection{Eve's information}

To calculate Eve's potential information we consider two strategies of classical communication between Alice and Bob during the post processing step: direct reconciliation and reverse reconciliation. In the first strategy Alice sends correcting information to Bob, and in the second one Bob sends it to Alice. 
We also assume, that after the beam splitting Eve is not restricted to any practical way of information extraction from this state, thus her potential knowledge is bounded by the Holevo information \cite{Holevo:73}. In the general case, the Holevo information $\chi$ sets the upper bound on the information which can be transmitted by a state randomly chosen from a set of $N$ states $\hat\rho_k$ with a respective probability $p_k$:
\begin{equation}\label{Holevo}
	\chi=S\left(\sum\limits_{k=1}^{N}p_k\hat\rho_k\right)-\sum\limits_{k=1}^{N} p_k S(\hat\rho_k),
\end{equation}
where $S(\hat\rho)$ is the von Neumann entropy $S(\hat\rho)=-{\rm Tr}\hat\rho\log_2\hat\rho$.

\subsubsection{Direct reconciliation}

In the direct reconciliation case, Eve has a state (\ref{bestates}) conditioned only on Alice's sent state $\ket{\alpha_k}$. Eve's conditional state is pure, thus her information is equal to the von Neumann entropy $I_{AE}=S(\hat\rho_E)$ of her unconditional state $\hat\rho_E=\frac{1}{N}\sum\limits_{k=1}^{N}\hat\rho_k$.

To calculate $S(\hat\rho_E)$ we need to find the eigenvalues of $\hat\rho_E$. The rotational symmetry of the phase--shift alphabet allows us to write Eve's conditional states in an orthogonal basis $\{\ket{m}\}$ as \cite{Chefles:98}:
\begin{equation}\label{Evestates}
	\ket{\epsilon_k}=\sum\limits_{m=1}^{N}c_m e^{i\frac{2\pi}{N}km}\ket{m}.
\end{equation}

In the basis $\{\ket{m}\}$ Eve's unconditional state takes the diagonal form:
\begin{equation}
 \hat\rho_E=\frac{1}{N}\sum\limits_{k=1}^{N}\ket{\epsilon_k}\bra{\epsilon_k}=\sum\limits_{m=1}^{N}|c_m|^2\ket{m}\bra{m},
\end{equation}
so Eve's information is equal to $I_{AE}=S(\hat\rho_E)=H[C]$,
where $H[C]$ is the Shannon entropy $(\ref{ShanEnt})$ of the probability distribution $C=\{|c_1|^2,|c_2|^2,\ldots,|c_N|^2\}$.

The coefficients $|c_m|^2$ are derived from a system of $N$ linear equations enumerated by an index $k=1\ldots N$:
\begin{equation}\label{Cmatr}
	\sum\limits_{m=1}^{N}e^{i\frac{2\pi}{N}km}|c_m|^2=\braket{\epsilon_N}{\epsilon_k}.
\end{equation}
It has a formal analytical solution 
\begin{equation}
|c_m|^2=\frac{1}{N}\sum\limits_{n=1}^{N}e^{-i\frac{2\pi}{N}mn-a^2(1-\eta)\left(1-e^{i\frac{2\pi}{N}n}\right),}
\end{equation}
where the coefficients $c_{m}$ depend on the signal amplitude $a$ and the channel transmittance $\eta$.

\subsubsection{Reverse reconciliation}

In the case of reverse reconciliation Eve has a state (\ref{bestates}) conditioned on Bob's measured state $\ket{\beta}$. After classical communication Eve can possibly find out the amount of information (\ref{Iab}) between Alice and Bob in each transmission, so we assume this value is publicly open. Additionally we assume that Bob announces the amplitude of the measured state, so Eve knows the measured state $\ket{\beta}$ up to a cyclic phase shift $2\pi/N$. We denote these possible states as $\ket{\beta^{(l)}}$. On her side, Eve has to distinguish between the states $\hat\rho_E^{(l)}=\sum\limits_k p_k(\beta^{(l)})\ket{\epsilon_k}\bra{\epsilon_k}$.

After averaging, Eve's state is $\frac{1}{N}\sum\limits_l\hat\rho_E^{(l)}=\hat\rho_E$, so the left entropy term in (\ref{Holevo}) is the same as we calculated before for the case of direct reconciliation. Due to the $2\pi/N$ phase--shift symmetry of the alphabet, the averaging in the right entropy term in (\ref{Holevo}) is just equal to the entropy of any of the states  $\hat\rho_E^{(l)}$, let it be the first one $\hat\rho_E^{(1)}$.

Again, we can rewrite Eve's states $\ket{\epsilon_k}$ in the orthogonal basis (\ref{Evestates}). Unfortunately, the state $\hat\rho_E^{(1)}$ in this basis takes a non--diagonal form $\hat\rho_E^{(1)}=\sum\limits_{k,m,n}p_k(\beta^{(1)})c_{m}c^*_{n}e^{i\frac{2\pi}{N}k(m-n)}\ket{m}\bra{n}$. We analytically calculate eigenvalues of this state for a given $N$, but the result is too large to be presented here. Finally, Eve's information is $I_{BE}(\beta)=S[\hat\rho_E]-S[\hat\rho_{E}^{(1)}]$.

\subsection{Postselection}

The key rate $G$, i.e. the amount of secret information per transmission (\ref{ABflow}), is equal to the difference between Bob's and Eve's informations \cite{Devetak:05,Renner:07}:
\begin{equation}\label{G}
	G=\int p(\beta)G(\beta)d\beta,\quad G(\beta)=I_{AB}-I_{AE,BE}.
\end{equation}
where $I_{AE}$ and $I_{BE}$ refer to direct and reverse reconciliation respectively. 

\begin{figure}[h]
\begin{center}
\includegraphics[width=6cm]{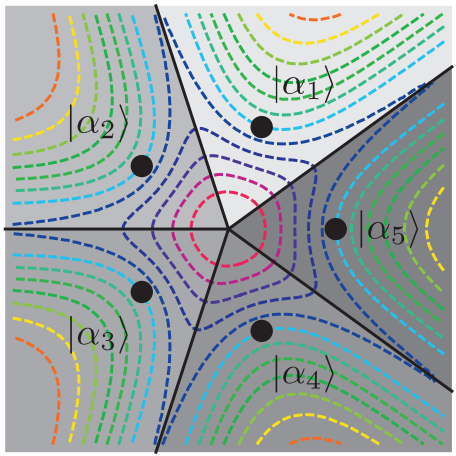}
\caption{Reconciliation and postselection areas for 5 letter protocol. Different shades of grey correspond to the regions in the phase space where measurement results $\ket{\beta}$ are associated with a given letter. Letters are shown as black circles. Dashed lines show the borders of the postselection areas for a case when amplitude is $a=1.4$, and transmittance varies from $0.95$ (the smallest area) to $0.4$ (the biggest area) with a step $0.05$.}
\label{5lett}
\end{center}
\end{figure}

In the case of direct reconciliation, Eve has to guess what was sent by Alice. If the channel transmittance is lower than $50\%$, Eve can potentially have a better signal than Bob, thus Eve's information can be higher than Bob's information and no secure communication is possible. To overcome this ``3 dB limit'' we use the postselection idea \cite{Silb:02}, so that Bob has an information advantage over Eve ($I_{AB}>I_{AE}$), i.e. we select only that part of transmissions which give us positive terms $G(\beta)$.

The postselection procedure in the direct reconciliation scenario can be qualitatively described as follows: Eve's information $I_{AE}$ does not depend on Bob's measured state $\beta$, so the key rate can be increased if Bob accepts only those transmissions where $\beta$ is such that he has higher information than Eve ($I_{AB}(\beta)>I_{AE}$). Instead of integration over the whole phase space (\ref{G}) we have integration over the postselected area (PSA):
\begin{equation}\label{Gps}
	G_{PS}=\int\limits_{PSA} p(\beta)\left(I_{AB}(\beta)-I_{AE}\right)d\beta.
\end{equation}

To find this PSA we numerically solve an equation $I_{AB}(\beta)>I_{AE}$. As an example in Fig.~\ref{5lett} we show the borders of the PSA for 5 letter protocol as the dashed lines. Different dashed lines correspond to different values of transmittance $\eta=0.4, 0.45, 0.5, \ldots, 0.95$, and the amplitude is fixed $a=1.4$.  The PSA is the phase space except the central region bounded by a dashed line. If the measured state $\ket\beta$ lies inside the region the transmission should be omitted, otherwise it is accepted. The higher the transmittance, the smaller the omitted region, the bigger the PSA, and the higher the key rate (\ref{Gps}).

To find the key rate $G_{PS}$ as a function of transmittance $\eta$ we optimize the amplitude $a$ such as to maximize the key rate $G_{PS}(\eta)=\max\limits_{a} G_{PS}(\eta,a)$. 

In the case of reverse reconciliation, Eve has to guess Bob's measurement result, thus her information cannot be higher than Bob's information: $I_{AB}\geq I_{BE}$. Therefore, $G(\beta)$ is always nonnegative, and the postselection procedure does not have to be applied.

\section{Results}

\begin{figure}[h]
\begin{center}
\includegraphics[width=12cm]{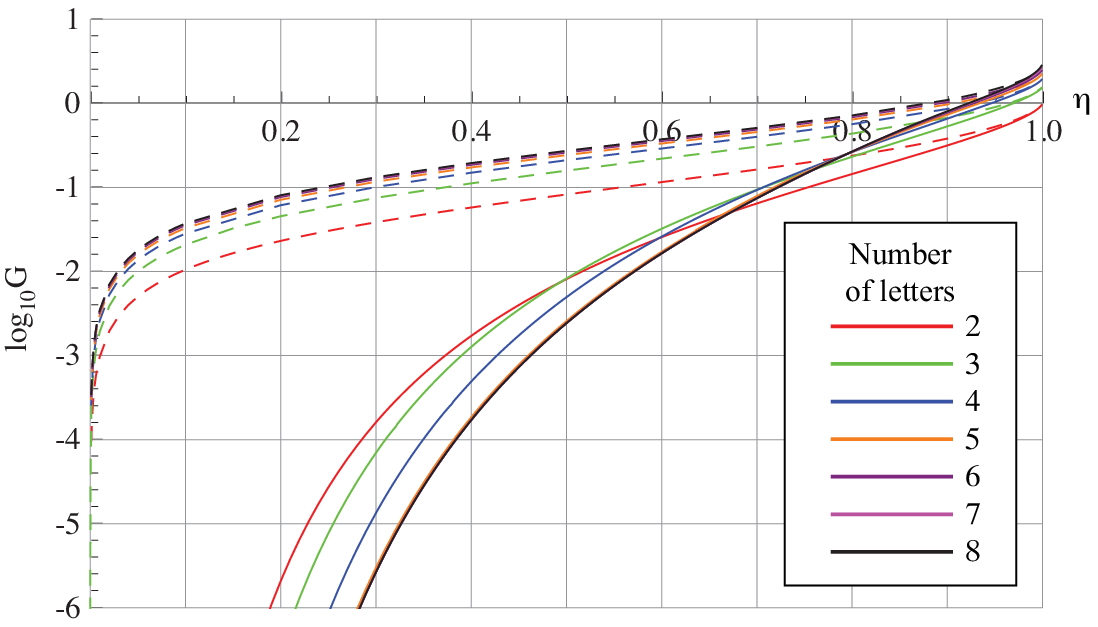}
\includegraphics[width=12cm]{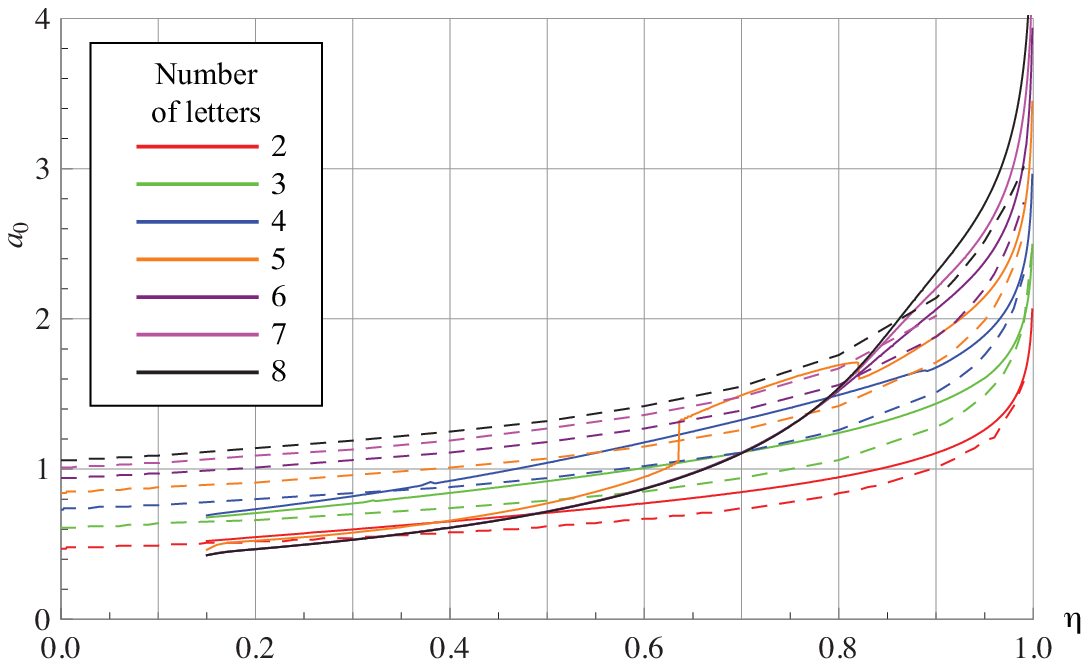}
\caption {The secret key rate $G$ in logarithmic scale (upper plot) and optimal signal amplitude (bottom plot). Solid and dashed lines correspond to direct and reverse reconciliation, respectively.}
\label{Gaeta}
\end{center}
\end{figure}

The calculated secret key rate $G_{PS}(\eta)$ and the optimal amplitude $a_0(\eta)$ for several alphabets are shown in Fig.~\ref{Gaeta}\footnote{Discontinuity of the curve $a_0(\eta)$ for the 5--letter alphabet is not a mistake. Due to the fact that the function $G(a)$ at a fixed value $\eta$ can have two slightly different global maxima, the exact optimization for variable $\eta$ may cause a ``jump'' from one maximum to another.}. We can see, that for the channel transmittance $\eta<0.9$ all the curves of the key rate are almost the same when the number of letters is more than 4. 

In the case of reverse reconciliation we have an interesting result: the higher the number of letters, the higher the key rate. We don't have an analytical expression for the $\infty$--letter alphabet, but its approximation by a high number of letter confirms that this is the best choice for all values of $\eta$. The key rate is higher than for the 2--letter alphabet of almost an order of magnitude.

In the case of direct reconciliation, we can see that lines $G(\eta)$ for various numbers of letters are intersecting in different points, which are presented in Table~\ref{etatab}. This means that for different values of transmittance $\eta$ there are different optimal numbers of letters. The higher the transmittance, the higher the optimal number of states. 
\begin{table}[ht]
\caption{Values of transmittance $\eta$, where a curve $G_{PS}(N,\eta)$ intersects with a curve $G_{PS}(N+1,\eta)$.}
\begin{center}
\begin{tabular}{ c c c c c c c c }
\hline
\hline
N & 2 & 3 & 4 & 5 & 6 & 7 & 8 \\
\hline
$\eta$ & 0.493 & 0.705 & 0.797 & 0.797 & 0.753 & 0.725 & 0.696\\
\hline
\hline
\end{tabular}
\end{center}
\label{etatab}
\end{table}

Again, we don't have have an analytical expression for the curve $G(N=\infty,\eta)$, but we found that left side of the curves quickly saturates (there is no essential difference between $G(N=5,\eta)$ and $G(N=64,\eta)$ for $\eta<0.9$). So we can conjecture, that the optimal alphabets can consist of 2, 3, 4, or $\infty$ letters. A curve $G(N=\infty,\eta)$ intersects with $G(N=4,\eta)$ at the value $\eta\simeq0.795$, so the most significant advantage of the multi letter protocol over the two letter protocol one can get in the case of high transmittance. The intuitive explanation is as follows:

In the case of high losses, Eve has a stronger signal than Bob. Thus the amplitude of the signal $a_0$ must be small, and Bob relies basically on the postselection. In postselection it is harder for Bob to distinguish between many letters than between two. Thus with an increasing number of letters his information essentially decreases. In this case the alphabet with two letters outperforms the multi letter alphabet.

In the opposite case of low losses Eve has a weaker signal, and Alice can increase the amplitude of the signal and number of letters. With higher amplitude of the signal Bob can better distinguish between many letters and increase his information. In the limit $\eta\rightarrow 1$ Alice can use signals with high amplitude and Bob can get almost $\log_2 N$ bit per transmission. Therefore, the more letters in the alphabet are, the higher Bob's information is. In principle, one can use an arbitrary high number of letters, and in the limit $N\rightarrow\infty$ (continuous phase modulation) Bob's information seems to be infinite. However, there are limiting factors from both experimental and theoretical viewpoints. First, as one can see in Fig.~\ref{Gaeta} the curves $G(\eta)$ and $a_0(\eta)$ start to essentially increase from the values $\eta>0.99$. In any real experimental setup there are imperfections (inaccuracy, losses, etc.), so the case $\eta>0.99$ can hardly be achieved. Second, any real signal has certain energy limit, which sets maximum amplitude. Also taking into account excess noise in the channel might somewhat change the situation.

\section{Conclusions}

To summarize, we have presented a new CV QKD protocol with coherent states. The protocol employs multi letter phase--shift--keying and heterodyne measurement. Security analysis of the proposed protocol is performed for the case of lossy but noiseless quantum channels. We have shown that for each given channel transmittance one can find a certain optimal number of letters (2, 3, 4, or $\infty$), optimal amplitude of the signal (typically, 1 to 4 photons per pulse), and optimal postselection threshold, which increase the secret key rate about one order of magnitude comparing to the protocol with binary modulation.

\section*{Acknowledgments}
The authors thank Norbert L\"utkenhaus for helpful discussions, Dominique Elser and Christoffer Wittmann for valuable comments on the manuscript. D.S. acknowledges the Alexander von Humboldt Foundation for a fellowship.

\section*{Bibliography}
\bibliographystyle{JphysicsB}


\end{document}